
\magnification=\magstep1
\hsize=16.0  true cm
\baselineskip=20pt
\parskip=0.2cm
\parindent=1cm
\raggedbottom

\def\pp{\parshape 2 0truecm 16truecm 1truecm 15truecm}
%
\def\apjref#1;#2;#3;#4 {\par\pp#1,  #2,  #3, #4 \par}
%
%
\def\oldapjref#1;#2;#3;#4 {\par\pp#1, {\it #2}, {\bf #3}, #4. \par}
%
%
\def\apj{ApJ}
\def\aap{A\&A}
\def\aaps{A\&AS}
\def\mnras{MNRAS}
\def\aj{AJ}
\def\araa{ARA\&A}
\def\pasp{PASP}
\def\apjs{ApJS}
\def\header#1{\bigskip\goodbreak\noindent{\bf #1}}
%
%
\def\subsection#1{\goodbreak\noindent\underbar{#1}}
\def\subsubsection#1{{\noindent \it #1}}
\def\ltsima{$\; \buildrel < \over \sim \;$}
\def\simlt{\lower.5ex\hbox{\ltsima}}
\def\gtsima{$\; \buildrel > \over \sim \;$}
\def\simgt{\lower.5ex\hbox{\gtsima}}
\def\etal{{\it et al.\ }}
\def\cf{{\it cf.\ }}
\def\eg{{\it e.g.\ }}

\def\lta{\mathrel{\spose{\lower 3pt\hbox{$\mathchar"218$}}
     \raise 2.0pt\hbox{$\mathchar"13C$}}}
\def\gta{\mathrel{\spose{\lower 3pt\hbox{$\mathchar"218$}}
     \raise 2.0pt\hbox{$\mathchar"13E$}}}
\magnification=\magstep1
\baselineskip=21pt
\def\aap{A\&A}

\def\aaps{A\&AS}
\def\pasp{PASP}
\def\kms{{\rm km s^{-1}} }
\centerline {{\bf CHEMISTRY AND KINEMATICS IN THE SOLAR
NEIGHBORHOOD:}}
\smallskip
\centerline {\bf {IMPLICATIONS FOR STELLAR POPULATIONS AND FOR GALAXY
EVOLUTION} }
\bigskip

\bigskip

\centerline {ROSEMARY F.G.~WYSE}
\centerline {Department of Physics and
Astronomy\footnote{$^\spadesuit$}{Permanent address}}
\centerline {The Johns Hopkins University}
\centerline {Baltimore, MD 21218}
\centerline {USA}
\medskip
\centerline {Institute of Astronomy}
\centerline {Madingley Road, Cambridge CB3 0HA}
\centerline {England, UK}
\centerline {and}
\medskip
\centerline {Center for Particle Astrophysics}
\centerline {University of California}
\centerline {Berkeley, CA 94720}
\centerline {USA}

\bigskip
\centerline {GERARD GILMORE}
\centerline {Institute of Astronomy}
\centerline {Madingley Road, Cambridge CB3 0HA}
\centerline {England, UK}

\vfill\eject

\centerline{ABSTRACT}

\noindent
The immediate Solar neighborhood should be a fair sample of the local
Galaxy. However, the chemical abundance distribution of long-lived
disk stars very near the Sun contains a factor of five to ten more
metal-poor stars, $-1 \simlt {\rm [Fe/H]} \simlt -0.4$ dex, than is
consistent with modern star-count models of larger scale Galactic
structure.  The metallicity distribution of complete samples of
long-lived stars has long been recognised as providing unique
constraints on the early stages of chemical evolution of the Galaxy,
so that one would like to resolve this anomaly.  We present a new
derivation of the local G-dwarf metallicity distribution, based on the
Third Gliese catalog combined with Olsen's (1983) Str\"omgren
photometry.  Kinematic data for these same stars, as well as for a
high-precision sample studied by Edvardsson {\sl et al.}  (1993),
provide clear evidence that the abundance distribution below
[Fe/H]$\sim -0.4$ contains two over-lapping distributions, the thick
disk and the thin disk. However these samples in isolation do not
allow a reliable deconvolution of the relative numbers in each
population. We achieve this by comparing the local metallicity
distribution with a recent determination (Gilmore, Wyse \& Jones 1995)
of the metallicity distribution of stars, selected with the same
evolutionary criteria as applied to our nearby star sample, but found
{\sl in situ} some 1500pc from the Sun. The gravitational sieve of the
Galactic potential acts on this second sample to segregate the low
velocity dispersion, thin-disk, component of the local sample, leaving
predominantly the second, higher velocity dispersion component.  Thus
the two samples are complementary, with the local sample providing
accurate data primarily for the thin disk, but weak
thin-disk/thick-disk discrimination, and the distant sample providing
excellent thick-disk/thin-disk segregation. We are thus able to break
the degeneracy between overlapping phase-space distributions.  That
is, the combination of these two datasets allows us to determine the
source of the local paradox: there is a substantial tail of the thin
disk (defined kinematically) metallicity distribution, which extends
below ${\rm [Fe/H] \approx -0.4}$dex. This is a robust conclusion,
being consistent with the sum of star count, stellar spatial density
distribution, and both local and distant kinematic and chemical
abundance data.  Using the sum of this information, we
deconvolve the local abundance distribution
function into thin disk and thick disk components, and show these
overlap considerably. This overlap has  implications for some
dynamical merger models of the formation of the thick disk.
The observed scatter in the
thin disk age-metallicity relation however obscures any simple
interpretation in terms of thick disk formation models.
The chemical abundance distributions derived in this paper, which are
based on the sum of local and distant data, and so are directly
integrated over the Solar Cylinder, improve long-standing constraints
on Galactic disk evolution. When combined with age and element ratio data,
tight constraints on the evolution of the disk will be available.

\vfill\eject
\bigskip
\centerline {1. INTRODUCTION}

The metallicity distribution of complete samples of long-lived stars
has long been recognised as providing unique constraints on the early
stages of chemical evolution of the Galaxy.  The main sequence
lifetime of F/G dwarf stars can be greater than the age of the Galaxy
and hence such stars provide a complete record of the chemical
evolutionary history (\cf\ van den Bergh 1962; Schmidt 1963; Pagel and
Patchett 1975).  Pioneering studies focussed on the only
reasonably-complete sample available, which is that for stars in the
immediate solar neighborhood; in effect stars within about 20pc of the
Sun.  Until very recently, kinematic data were considered only as a
means to quantify, and to correct for, biases in the sample used to
derive the chemical abundance distributions. It is now appreciated
however that the combination of chemical abundance and kinematic data
is a substantially more powerful determinant of Galactic evolution
than is either distribution considered in isolation. It is an improved
determination of that bivariate distribution function, and its
analysis, which is the theme of this paper.

The abundance data which have been available for the complete sample
of local stars, until now, have been derived from broad-band
photometric estimates, primarily using the `ultra-violet excess' (Sandage
1969), derived in practice from somewhat heterogeneous UBV photometric
data.  The `G-Dwarf Problem' was identified when these data were
compared with the predictions of the Simple, Closed-Box model of
chemical evolution, and has been characterised as `the deficit of
metal-poor stars in the solar neighborhood relative to the one-zone
model of Galactic chemical evolution' (Pagel 1989).

The Simple, Closed-Box model has the virtue of analytic simplicity,
and has the vice of being based upon physical assumptions that are
inappropriate for a disk galaxy. It is somewhat of a `straw man'
model, easily knocked down, involving as it does several stringent
assumptions.  The assumptions inherent in this model are that:

\item{i)}  the system under study has zero metallicity initially;
\item{ii)} the system is homogeneous at all times, {\it i.e.\/}
there is no intrinsic scatter in the chemical enrichment of the
interstellar medium,
and specifically a unique correspondence between time and metallicity;
\item{iii)} the system has a constant stellar Initial Mass Function (IMF);
\item{iv)} the system has constant total mass at all times, {\it
i.e.\/} flows of material do not occur;
\item{v)} the Instantaneous Recycling Approximation is valid for
nucleosynthesis products.

Lifting any (or all!) of these assumptions can provide for a very
different metallicity distribution for long-lived stars.  As we
discuss further below, gas flows and spatial and temporal
inhomogeneities in the interstellar medium are a natural feature of
real disk galaxies.  Even were the Simple Model predictions to fit
some observational dataset, its inherent implausibility means that it
is more likely that several compensating effects had generated this
agreement by chance, rather than that the underlying physical processes
corresponded to all the assumptions above.  The important strength of
the Simple Model lies in its ability to provide a convenient
parameterization of the extent to which its assumptions fail, rather
than in its ability to describe a dataset.

Chemical evolution models have been plagued with a myriad of
assumptions and parameters, which leads to a lack of uniqueness (Tosi
1988). In order to improve one's understanding of Galactic evolution,
one must consider the plausibility of the physics behind the model
assumptions, and subject the model to a range of observational tests.
Many of the simplest, single-parameter, adjustments to the Simple
Model can be excluded by considering the totality of relevant
data. For example, allowing the IMF to vary with time, as first
discussed by Truran and Cameron (1971) in the context of prompt
initial enrichment, is in conflict with the element ratio data which
are now available for metallicities spanning $-3$ dex to above solar
(Nissen \etal\ 1994).  Similarly, one cannot posit `pre-enrichment' as
a solution to the G-dwarf problem without identifying a plausible
source: stellar evolution in the stellar halo does not suffice
(Hartwick 1976).  It is clear, however, that detailed analysis of
local chemical abundance data does provide important constraints on
Galactic evolution, and is a worthwhile exercise.

The primary limitation of the nearby star sample utilised in studies
of the local chemical abundance distribution is its small size.  This
inevitably means that stars which are either intrinsically rare --
such as halo population subdwarfs -- or stars which are common in the
Galaxy as a whole, but whose spatial distribution and kinematics are
such that their local volume density is small -- such as thick disk
stars -- are poorly represented.  Most recent and current efforts to
extend present volume-limited samples to include minority populations
have, for practical observational reasons, utilized
kinematically-defined samples of stars observed in the solar
neighborhood.  Subsequent correction for the kinematic biases inherent
in these samples requires careful modelling (\cf\ Norris and Ryan
1991; Ryan and Norris 1993; Aguilar \etal\ 1995).  An {\it in situ\/}
sample, truly representative of the dominant stellar population far
from the Sun, is required before one can be confident that these
large, and model-dependent, corrections to local samples are reliable.

We have recently determined the (spectroscopic) iron abundance
distribution for a sample of F/G stars {\it in situ\/} 1--5 kpc from
the Galactic Plane (Gilmore, Wyse and Jones 1995).  These data provide
the volume-complete metallicity distribution at $z \simgt 1$ kpc which
is complementary to local samples, and which allows a quantitative
assessment of the reliability of corrections to solar neighborhood
samples to account for their inherent kinematics biases, and of the
statistical validity of the local sample.  Combined with the local
sample, these new data provide a powerful test of the integrated
chemical evolution and star formation history of the local Milky Way.

As discussed below, recent determinations of the Solar neighborhood
abundance distribution have isolated (but not noted) an anomaly: while
the derived distribution of chemical abundances for long-lived stars
in a column through the Galactic disk at the Sun has reinforced the
reality of the deficit of metal-poor stars, relative to Simple Model
expectations, it has also derived a fraction of metal-poor stars which
is a factor of nearly an order of magnitude too high for consistency
with star-count models which explain data for distant stars. Typical
star-count models derive a local ratio of thick disk to thin disk
stars which is a few percent (e.g. reviews by Gilmore, Wyse and
Kuijken 1989; Gilmore 1990; Majewski 1993). Kinematic studies (Carney,
Latham and Laird 1989; Freeman 1991; Edvardsson \etal\ 1993) suggest
the thick disk becomes apparent at abundances below ${\rm [Fe/H]
\approx -0.4}.$  As discussed below,
recent analyses of local chemical evolution, in
contrast, derive a fraction of typically 25 percent of `disk' stars
with ${\rm [Fe/H] \leq -0.4}$ (Pagel 1989; Sommer-Larsen 1991).  Thus,
the local abundance distribution for disk stars apparently contains
too many `thick disk' stars to be consistent with other,
well-established, analyses of Galactic structure. We quantify and
resolve this discrepancy in this study.

The present paper builds upon  previous work deciphering the chemical
evolution of the Galactic disk.  We provide a new
determination of the solar-neighborhood metallicity distribution,
derived from high-precision, intermediate-band photometry of
long-lived stars.  We compare this distribution with our data for
distant F/G dwarfs, scaled down to the Galactic plane, and combine
these two datasets.  We use the joint distribution to derive the
abundance distribution of metal-poor thin disk stars, showing this is
consistent with the distribution required by star count models. We
then deduce the constraints on the formation of the thick disk, and the
early evolution of the thin disk, which follow from this deconvolution.

\vfill\eject
\centerline {2. THE SOLAR NEIGHBORHOOD G-DWARF METALLICITY DISTRIBUTION}
\smallskip

\centerline {\it 2.1 Previous Determinations}

Previous analyses have shown the power, and limitations, of local
samples of F/G dwarfs in constraining the chemical evolution of the
disk.  The sample of Pagel and Patchett (1975) is the most often
utilised.  This consists of G-stars drawn from various catalogs, the
complete sample (to distances of 25pc) being 133 stars from the RGO
catalog of nearby stars, with metallicities derived from UV excess.
The derived metallicity distribution of the `solar cylinder' -- that
volume of unit circular cross-section centered on the Sun's position
and extended vertically through the extent of the disk -- is taken
from their Figure 6 and shown here in Figure 1(a).  Note that only the
binned data were published by Pagel and Patchett, and that those
presented here are uncorrected for observational error and/or
intrinsic scatter.  The use of UV excess provides only rather
inaccurate metallicity estimates (denoted [Fe/H]$_\delta$); Norris and
Ryan (1989) derive a typical error of 0.45 dex for metallicities
estimated with this technique, at least for their sample of interest,
with [Fe/H]$_\delta < -0.4$ dex. Thus, the resulting abundance
distribution will be substantially broadened, by measurement
uncertainties, relative to the true underlying distribution. Since the
width of the predicted distribution function is one of the
characteristic differences between the simple model and more realistic
models, such as those involving flows (\eg\ Edmunds 1990) or
inhomogeneities even in a closed system (\eg\ Tinsley 1975), the
precision of the abundance calibration is of major importance.  Pagel
and Patchett adopted a broadening of only 0.2 dex for the combination
of measurement uncertainty and intrinsic scatter.

In an extension of his earlier work, Pagel (1989) adopted a
re-calibration of the relationship between [Fe/H] and UV excess due to
Cameron (1985).  From the same photometric data as his earlier work,
he derived the metallicity distribution shown in Figure 1(b).  In
addition to the considerable random uncertainties from the UV excess
abundance calibration, Cameron's calibration is systematically $\simlt
0.2$ dex more metal-poor in [Fe/H], for given UV excess, than the more
widely-used calibration of Carney (1979).  Despite the large scatter
referred to above, Norris and Ryan (1989) found no {\it systematic\/}
offset between [Fe/H]$_\delta$ based on the Carney calibration and
their spectroscopically-determined metallicities.  Again, their sample
was limited to stars with [Fe/H]$_\delta < -0.4$ dex, but it is just
these stars which are of greatest interest for the solar cylinder
G-dwarf metallicity distribution.

The metallicity distributions reproduced in Figures 1a,b are not in
fact those obtained directly from the photometric data for the local
sample, but are for the `solar cylinder' and have been weighted (by
Pagel) to attempt to take account of the kinematic bias introduced by
the solar neighborhood selection -- high velocity stars, which spend
only a small fraction of their orbit in the vicinity of the Sun will
be under-represented.  In the simplest picture of disk structure and
evolution there are no radial flows, or gradients in either density or
kinematics, and an integrated history of the disk is obtained by study
of the stars in unit area of the plane, rather than unit volume.  If
density profile data were available, then one could simply integrate.
However, such data were not available until the mid 1980s, so Pagel
(following Schmidt 1963) weighted the input binned metallicities by
the mean vertical velocity of stars in that bin.  However, this
correction is only appropriate for motion in a harmonic potential
(constant period for all orbits), which corresponds to a constant
density distribution of gravitating material. While this may be a
reasonable approximation for low-velocity stars that do not penetrate
much beyond a disk scale height, it is a rather poor approximation for
just those high-velocity stars for which the weighting is most
important.  In particular, the thick disk and halo remain
under-represented (\cf\ Gilmore and Wyse 1986; Gilmore 1990).

Sommer-Larsen (1991; see also Sommer-Larsen and Antonuccio-Delugo
1993) followed Pagel's (1989) sample selection and UV-excess
calibration, and so analysed the same basic metallicity distribution.
Again, they followed
Pagel (1989) in
recognising that the photometrically-determined
metallicity correlated most strongly with the {\it iron\/} abundance,
which is perhaps the element least likely to be well-described by the
Instantaneous Recycling Approximation,  inherent in the
chemical evolution models that are most often compared with the
metallicity distributions, given that Type Ia supernovae are an
important site for iron synthesis.  Rather than calculating
predictions from more realistic chemical evolution models,
Sommer-Larsen instead adopted a mean relationship between oxygen,
produced predominately in massive stars by Type II supernovae, and
iron, and transformed the photometric metallicity/iron distributions
into `oxygen' distributions.  It is now clear from the Edvardsson {\it
et al.\/} (1993) data that the relationship between oxygen and iron
varies as a function of radius in the Galaxy, as expected if the star
formation rate in the galactic disk varies as a function of radius,
being higher in the central regions (cf. Wyse 1995).
The adoption of a single-valued
relationship introduces uncertainties of perhaps 0.2 dex for the more
metal-poor stars in the sample.  We here have reverse-transformed the
oxygen abundance distributions of Sommer-Larsen (1991) back to
photometric-metallicity distributions.  Sommer-Larsen (1991) improved
upon the harmonic potential weighting by calculating the potential of
a self-gravitating disk that had the observed stellar-density profile
of Kuijken and Gilmore (1989).  This provides more weight to the
`hotter' metal-poor stars, and results in the metallicity distribution
shown in Figure 1(c).

We return to discussion of the appropriate distribution of chemical
abundances to describe the Solar neighborhood below.  However, we note
here that the fractions of stars more metal poor than $-0.4$ dex,
 where thick disk kinematics are apparent (see below),   in
these weighted distributions are approximately 25\% (Pagel and Patchett),
40\% (Pagel), and 50\% (Sommer-Larsen).  Were this to be taken as an
estimate of the relative mass of the thick disk to that of the thin
disk, one would conclude that the thick disk is of order the same mass
as the thin disk.  As we discuss below, this would be somewhat
difficult to understand in the light of star count analyses and {\it
in situ\/} studies of the thick disk.

\bigskip
\centerline {\it 2.2 A New Determination -- Sample Selection}

Photometric catalogs of nearby stars have progressed substantially
in both precision and reliability in the almost 20 years since Pagel
and Patchett obtained their chemical abundance distribution function.
Intermediate-band Str\"omgren photometry provides a more accurate
estimation of the metallicity of a F/G star than does the broad-band
color-based UV-excess, and is now available for many nearby stars.
The sample of long-lived, main-sequence stars in the immediate solar
neighborhood, for which carefully checked data are given in  the current
version of the Gliese catalog, has been extended from the distance limit of
about 20pc, available at the time of the Pagel and Patchett study,
 to distances of about 30pc.  As such, the Gliese catalog
provides  the ideal local sample,
superseding the RGO catalog. Hence
we combined the most reliable census of
nearby stars with the most precise
photometric data, by cross-correlating the Third Gliese catalog
(Nearby Stars 3, obtained from NSSDCA) with the $uvby(\beta)$
photometry catalog of Olsen (1983; also obtained from NSSDCA).  Both
catalogs were searched to identify all single, normal stars (no
white dwarfs) with ${\rm V \le 9; \,\,\, 0.4 \le B-V < 0.9}$, which
appeared in both.

This produced 128 stars; one which is known to have an erroneous
parallax tabulated was immediately deleted (HD 140283; see Gilmore,
Edvardsson and Nissen 1991), leaving a sample of 127 F/G dwarfs.
For comparison, the sample of Pagel and Patchett, and later authors,
contained 133 stars with data from heterogeneous sources.
\bigskip
\vfill\eject
\centerline {\it 2.3 Metallicity Estimator}

The most precise and reliable metallicity indicator which is available
for essentially the whole sample of nearby stars is the
relationship between
[Fe/H] and Str\"omgren photometric index $(b-y)$.
We have adopted
the calibration of this photometric metallicity
estimator  due to Schuster and Nissen (1989a).
This estimator, [Me/H], is given by the following
expression for stars with $0.22 \le (b-y) \le 0.375:$ $$\eqalign {{\rm
[Me/H]} =1.052&-73.21m_1+280.9m_1(b-y)+333.95m_1^2(b-y)
-595.5m_1(b-y)^2 \cr &+[5.486-41.62m_1-7.963(b-y)]\log(m_1-c_3),\cr}
$$ where $c_3=0.6322-3.58(b-y)+5.20(b-y)^2$,

\noindent and by the following
when $  0.375 < (b-y) \le 0.59: $
$$\eqalign {  {\rm       [Me/H]} =&-2.0965+22.45m_1-53.8m_1^2-62.04m_1(b-y)\cr
           & +145.5m_1^2(b-y)
       +[85.1m_1-13.8c_1-137.2m_1^2]c_1.\cr}
$$
Note that this calibration
does not require a measurement of $\beta$,
the Str\"omgren index that is most sensitive to effective temperature.
The $\beta$ index, being narrow band rather than intermediate band,
is available for fewer stars than are $uvby$ data.

Edvardsson \etal\ (1993) provide a metallicity calibration for F/G
stars that does utilise $\beta$, and it is possible to make a
direct  comparison between
the two techniques.  The
Edvardsson \etal\ estimator is $$ {\rm [Me/H]} = -[10.5+50(\beta -
2.626)] \times \delta m_1 +0.12$$ where $\delta m_1 = m_{1,Hyades}
(\beta) -m_1$, with the Hyades sequence taken from Crawford and Barnes
(1969); note that their relationship between $m_1$ and $\beta$ for the
Hyades is well-behaved for $2.6 \le \beta \le 2.89$ only.

 The metallicity estimates obtained using these two
techniques, for the ten stars in the present sample with $\beta$ in
the well-behaved Hyades range, and with
 $(b-y)$
within  the Hyades  range of 0.076
-- 0.39 (Crawford and Barnes 1969) are shown in Figure 2.
The scatter about the dotted line which indicates perfect
correspondence is less than 0.1 dex.  Indeed,  the
mean difference between the two techniques for this subsample
is 0.04 dex, with a
dispersion of 0.05 dex.  Thus adopting the Schuster and Nissen
metallicity calibration is justified.

Further, Edvardsson \etal\ investigated the calibration of their
photometric metallicity estimate by comparison with their
spectroscopic iron
abundance estimates. They found, after removal of  15  of their most
metal-rich stars (all with metallicity above the solar value),
very good agreement: $${\rm [Me/H] = 1.049[Fe/H]}
-0.030; \qquad \sigma = 0.074 \qquad n=174.$$ This result is
valid for their entire range of iron abundances, including the
bulk of their metal-rich stars.
Without removal of those 15 metal-rich
stars, for which the photometric indices yield an overestimate of
the iron abundance by 0.1 -- 0.3 dex, it can be seen from their
Figure 6 that the scatter increases to $\sim 0.15$ dex above
solar metallicity, with a mean offset of $\sim 0.05$ dex.

There are four stars in the present sample that have
spectroscopic iron abundances available from Edvardsson \etal\
These are HD 6434, with a spectroscopic ${\rm  [Fe/H] = -0.54}$ dex,
HD 22879, with a spectroscopic ${\rm [Fe/H] = -0.84}$ dex,   HD 89707,
with a spectroscopic ${\rm [Fe/H] = -0.42}$ dex and   HD 165401, with a
spectroscopic ${\rm [Fe/H] = -0.47}$ dex.  Comparison of our metallicity
estimates with the iron abundances yields a mean offset of
$0.003$ dex (their estimate being the one slightly more metal
poor) and a dispersion of only 0.04 dex.  Thus the metallicity
estimate adopted here based on intermediate-band photometry is a
reliable and precise ($\sigma \simlt 0.1$ dex) indicator of {\it
true iron abundance}.

Applying the Schuster and Nissen calibration to the F/G star
sample extracted from the Gliese catalog provides the metallicity
estimates given in Table 1, which also contains the necessary
photometric data for the stars.  The addition of the Sun brings
the sample to 128 stars and results in the metallicity
distribution shown as a histogram in Figure 3.

\bigskip
\centerline {\it 2.4  Isolation of Potentially Long-Lived Stars}

In order for the metallicity distribution function to be an
unbiased description of the chemical history, the sample must be
restricted to include only those stars which have both masses and
metallicities such that any star ever formed, which has those
parameters, is still on the main sequence today.  If this
restriction were not applied, then some parts of age-metallicity
space would be excluded from measurement, potentially distorting
the resulting metallicity distribution function.  It should be
emphasized that the selection of a restricted range of spectral
types does {\it not\/} achieve this, due to metallicity-dependent
evolutionary times.  Spectral types in the absence of metallicity
data are not a sufficiently good indicator of mass, or of main-sequence
lifetimes.

An estimate of the age of the disk, based on (VandenBerg) stellar
evolution models, is available from the data of Edvardsson {\it et
al.} (1993).  This sample of 189 stars was drawn from the Olsen local
sample and was designed to cover the metallicity range above one-tenth
of solar uniformly; thus there should be no kinematic-metallicity
bias.  As discussed in some detail by Freeman (1991), the kinematic
thin disk/thick disk transition in that sample occurs fairly abruptly
at 12Gyr, which may be taken as a minimum age for the thin
disk.\footnote{$^\clubsuit$}{Note that the Edvardsson {\it et al.}
data show no evidence for an age gap between the thick disk and the
thin disk, either in the direct age estimates, or in the element
ratios (see Wyse (1995) for further discussion of this point).  The
data of Marquez and Schuster (1994) also show considerable overlap in
age between the halo and thick disk, with a continuous age distribution, into
the thin disk. } Thus the evolutionary selection was effected by
removing stars which had a potential main-sequence lifetime less than
12Gyr, according to the VandenBerg and Bell (1985) isochrones
(Y=0.20), the VandenBerg (1985) isochrones (Y=0.25) and the VandenBerg
and Laskarides (1987) isochrones.  Colors for models of solar
metallicity and above were initially obtained by interpolation in
Table 1 of VandenBerg and Bell (1985), using the effective temperature
and gravity of turnoff stars from VandenBerg (1985; solar metallicity)
and from VandenBerg and Laskarides (1987; three times solar
metallicity).  Interpolation in this Table predicts a solar color of
$(b-y)_\odot = 0.388$, to be compared with a measured color of
$(b-y)_\odot \simeq 0.405 \pm 0.002$ (Saxner and Hammarb\"ack 1985,
Magain 1987).  We took this difference in solar color as a zero-point
color offset, and thus adjusted the interpolated $(b-y)$ colors by
adding 0.017 ({\it cf.\/} Schuster and Nissen 1989b).  Note that there
is no agreement in the zero-point of the theoretical models in $c_1$
with the present data.  Fortunately for our purposes we need to use
only the $(b-y)$ color information, as only the turnoff temperature is
required and we have metallicity data.  The adjusted turnoff $(b-y)$
colors obtained this way are given in Table 2, for convenience.  These
models show an increasing, and probably over-estimated, sensitivity of
$(b-y)$ to metallicity, with increasing metallicity. In particular,
the large increase in $(b-y)$ based on the VandenBerg and Laskarides
models of above solar metallicity does not agree with the trend
predicted by the observationally-calibrated model atmospheres of
Edvardsson \etal\ (1993).  This latter study finds $\Delta
(b-y)/\Delta{\rm [Me/H]} = +0.026$ mag/dex, for stars with iron
abundance in the range from solar metallicity to +0.17 dex.  It should
be noted that since the VandenBerg (1985) isochrones have a 1M$_\odot$
star of solar metallicity at the turnoff after only 9Gyr, the 12Gyr
turnoff color for this metallicity should be at least as red as the
Sun, lending credibility to the interpolated value for solar
metallicity obtained here.  Accepting the solar metallicity 12 Gyr
turnoff parameters based on the 1985 VandenBerg models, then the
Edvardsson \etal\ trend predicts that the 12 Gyr turnoff color for a
population of metallicity equal to +0.5 dex should be $(b-y)_{+0.5} =
0.419$.  This is also given for convenience in Table 2.

Figure 4(a) shows the raw sample of Gliese F/G stars plotted in the
[${ c_1 - (b-y)}$] plane, together with the 12Gyr turnoff colors for
the relevant range of metallicities.  All stars blueward of the
turnoff for their metallicity (linear interpolation was used between
fiducial points in the turnoff-color--[Fe/H] relation) are excluded
from further consideration, since they occupy a region in which it is
possible that old stars would have evolved out of the sample.
Photometric errors are thus assumed to be unbiased.  It should be
noted that the Sun is excluded by this procedure; as noted above this
is not unexpected, given that the main-sequence lifetime of the Sun in
these models is only 9Gyr, which is less than our adopted age of the
disk. For the stars of metallicity greater than the solar value,
adopting the Edvardsson \etal\ turnoff color excludes only
three fewer stars than does use of the redder vandenBerg and
Laskarides turnoff color. In the following, we will use the results
from adoption of the Edvardsson \etal\ color, since it is
observationally-calibrated.

The excluded stars from the Gliese catalog, which total 36 (plus the
Sun), are indicated as such in Table 1 by a `yes' in the column headed
`rejected?'.  The sample remaining after this pruning, which totals 91
stars, has the color distribution shown in Figure 4(b).  The
metallicity histogram corresponding to this `G-dwarf' sample, is given
in Figure 5.  This is our final metallicity distribution of
long-lived stars observed in the solar neighborhood.  This
distribution will be combined with our metallicity distribution for
long-lived stars {\it in situ\/} several kiloparsecs below the
Galactic plane in the following sections.  Given that our main
scientific conclusions will be based upon comparing observations with
observations, rather than comparing observations with precise
theoretical predictions, we will not attempt any deconvolution of
observational uncertainties from our metallicity distributions.
Indeed, below we will rather re-bin the local sample to match the (larger)
uncertainties of the {\it in situ\/} data, $\sim 0.2$ dex.

 A comparison of the distributions in Figures 3 and 5 reveals that
stars more metal-rich than [Fe/H]$ \sim -0.4$ dex are preferentially
removed. A consistency check on the robustness of our rejection of
these stars may be made by comparison with two other samples.  The
first of these is the sample of around seven hundred K giants in the
solar neighborhood of McWilliam (1990); these stars, being evolved,
are by definition comparable to those stars rejected from our sample
of long-lived F/G dwarfs.  His sample contains every evolved G and K
star in the {\it Bright Star Catalog}, and is approximately
volume-complete, but with a bias to higher mass, more distant stars.
His abundance estimates are based on high-resolution spectra.  The
second sample consists of the F/G stars from the Edvardsson {\it et
al.} (1993) sample that are younger than 12 Gyr, our chosen cutoff
main sequence lifetime.  Edvardsson {\it et al.}  calculate the
corrections required to generate a volume-complete metallicity
distribution from their observed distributions.  The metallicity
distributions of these three samples -- those stars rejected by our
lifetime criterion (thus containing stars of all ages up to 12 Gyr),
the local G/K giants (McWilliam 1990) and the volume-corrected
Edvardsson {\it et al.} younger subsample -- are shown together in
Figure 6.  Note that we have chosen a wider bin size, 0.2 dex, to
ensure that any small zero-point or scale differences which may exist
between the different samples do not obscure the main point of the
comparison. It
is clear that the three distributions are all consistent with one
another, and provide a well-defined metallicity distribution function
for the younger thin disk.

It should also be noted that even the VandenBerg (adjusted) 12Gyr
turnoff for a stellar population of three times solar metallicity,
$(b-y) = 0.513$, is bluer than the red limit of the data, $(b-y) =
0.528$; this means that old, metal-rich stars have {\it not\/} been
excluded by our selection procedures.  This was not known to be the
case for previous samples.

\medskip\centerline {{\it 2.5  What Stars Have We Isolated?}}

The sample whose chemical abundance distribution we have obtained
will be a mix of all the stellar
populations that are represented in the solar
neighborhood -- effectively, the thick and thin disks.
Kinematics are usually invoked to determine the boundary of the thick
disk/thin disk transition.
As exemplified by the analysis of Freeman (1991; 1994), the
 vertical velocity distribution   of the Edvardsson \etal\ sample (which
has no kinematic-metallicity bias) can be
described by essentially two numbers -- ${\rm  <|W|>}$  approximately
constant at $19 \kms$ for thin-disk stars, and $ {\rm  <|W|> }  \simeq 42
\kms$ for thick-disk stars.  The value for the thin-disk
stars is generally understood as reflecting the saturation of the
disk-heating mechanism by scatterers confined to the disk plane (\eg\
Lacey 1991).  Freeman assigns age ranges to these kinematics, but one
can also look at metallicity ranges.  Following his approach, Figure 7
shows the cumulative ${\rm |W|}$ velocity against [Fe/H] rank for the
Edvardsson \etal\ sample; in such a plot a straight line has a slope
equal to the mean of the absolute value of the $W$
velocity;
this is an approximation to the dispersion if the underlying distribution
function is a single zero-mean Gaussian.

It is clearly consistent with the data to adopt the  colder
kinematics (indicated by the dashed line)
for the stars more
metal-rich than [Fe/H]$ \sim -0.4$
dex, whereas the hotter 
\vfill\eject
\noindent component
is evidently present in those stars with $-1.0 \le
{\rm [Fe/H] < -0.4}$.
Thus, purely kinematic definitions show the thick disk to be an
important component of the disk for stars with metallicity below
${\rm [Fe/H]} < -0.4$. We now check to see if it is a dominant or even
unique component.

Figure 5 above is the abundance distribution function
for our final local sample. Simple counting shows that the fraction of
the sample which has abundance below ${\rm [Fe/H]} < -0.4$ is $\simgt
25\%$, and below ${\rm [Fe/H]} < -0.5$ is $\simlt
20\%$. This can be compared with that derived from star count
models, which reproduce the number of stars {\it in situ} some 1--2
kpc from the Galactic plane, using a (highly correlated) combination
of a scale height and a local volume density, expressed as the ratio
of thick disk to thin disk stars. Typical star-count models provide
estimates of this local ratio of thick-disk to thin-disk stars of
between $\sim 2\%$ and $\sim 10\%$, with the most recent models being
in the middle of this range (Gilmore, Wyse and Kuijken 1989; Majewski
1993; Ojha {\etal} 1994 and references therein). That is, with
this metallicity identification of the thick disk, which is
apparently consistent with kinematic data, there is an apparent {\sl
overdensity} of thick disk stars in the local sample here by a factor of
about 5.  Thus there is a serious failing of at least one of the
star-count
models, the bivariate definition of the boundary between the
thick and the thin disks, or of the assumption that the thick disk --
thin disk boundary is discontinuous.

The star count models are sufficiently successful that the
possibility they are in error by the required magnitude
can be excluded   The bivariate
kinematics-chemistry determination of the {\sl onset} of the thick
disk is justified above. Thus we must deduce that there is a
substantial overlap between the thick disk and the thin disk.  That
is, the onset of the thick disk, as evidenced in the rapid change in
vertical velocity distribution (Figure 7)  at a metallicity
of $\rm [Fe/H] \sim -0.4$, does not correspond to the end of the thin
disk.
This interpretation requires that the vertical velocity
distribution function for these lower metallicity stars is not a
single Gaussian.

The composite nature of the velocity distribution is
confirmed in Figure 8a,b,c,d which show histograms of the
vertical velocity data for all the Edvardsson {\it et al.} stars
(panels (a) and (c)) and all the single Gliese F/G dwarfs (panels
(b) and (d)), with metallicity in the range $-0.4 \le {\rm
[Fe/H]} < -0.9$ (panels (a) and (b)) and with  metallicity in the
range $-0.5 \le {\rm [Fe/H]} < -0.9$ (panels (c) and (d))  (W
velocities for these stars are listed in Table 1; we note that
there is only one metal-poor star -- HD100004 -- that was
rejected from our potentially long-lived sample and which is in
this complete sample).  The smooth curve superposed corresponds
to a Gaussian with dispersion of 45 km/s; there is evidently a
cold core, reminiscent of the thin-disk kinematics, in addition
to the stars with hotter, thick-disk kinematics.  This cold
core persists even after the exclusion of stars in the range
$-0.4 \le {\rm
[Fe/H]} < -0.5$.  Note that we
are limited, by the small sample sizes, to this broad a
metallicity range, and so are unable from these data alone to
determine the true detailed metallicity distribution of local
stars with thin disk kinematics, but can say that it clearly
extends into the regime of the thick disk.  In contrast, stars
with thick disk kinematics are not a substantial contributor at
higher metallicities, ${\rm [Fe/H]}
\simgt -0.3$, or they would be visible in Figure 7.  We achieve
the deconvolution of the thick-disk  and thin-disk metallicity
distributions in section 3 below, by combining distant and local
data.

The fact that thin-disk stars are significant, if not dominant,
contributors to the metallicity distribution at ${\rm [Fe/H]} \sim
-0.6$, around the peak of the thick disk metallicity distribution,
results from the much higher local normalisation of thin disk stars.
As discussed by Gilmore and Wyse (1985), the old thin disk has a
characteristic value
in its metallicity distribution of ${\rm [Fe/H]} \sim -0.25$, and a
width which may be characterized crudely by a Gaussian of
dispersion $0.2$ dex. This combination of a higher local normalization
and a peak at a metallicity which is only as far as the spread of its
distribution from the maximum of the thick disk distribution, means
that a large part of the metal-poor local stellar populations are
members of the thin disk.
We return to this point in section 3.2 below.

\medskip
\centerline
{{ 3. THE SOLAR CYLINDER G-DWARF METALLICITY DISTRIBUTION}}

We have identified above a reliable and volume-complete, though small,
local sample of potentially long-lived stars. We recall that any such
sample in the Plane may under-sample high velocity dispersion
components of the disks, and so should be corrected accordingly.
Conversion of this sample into a fair estimator of the local
time--integrated chemical abundance distribution requires that this
sample be corrected from volume to surface density. The simplest
models of galaxy evolution assume there is no {\sl radial}
re-arrangement of chemically--enriched material, with vertical and
radial kinematics decoupled, so that one should integrate in a column
perpendicular to the disk. This may be achieved either by kinematic
weighting of a local sample, or by interpolating between a local and a
similar distant {\it in situ\/} sample. This second approach,
achieved for the first time in this paper, has the
additional benefit that the comparison may be used to estimate more
reliably than was possible above the relative contributions of low
velocity dispersion (thin disk) and high velocity dispersion (thick
disk) stars to the metal poor stellar distribution. We address both
options in this section.

\medskip\centerline {\it {3.1 Kinematic Weighting of the Local Sample}}

We now attempt to define a representative metallicity distribution
function from our local stellar sample using a kinematic-dependent
weighting.  This is to correct for the fact that the probability of
observing a star on a given orbit while it is in the solar
neighborhood depends on the fraction of the orbit that the star spends
there. It also provides for the transformation of volume densities to
surface densities, leading to the iron abundance distribution
integrated through the solar cylinder.  The correction factor is the
inverse of the fraction of an orbit that a star spends traversing the
disk, and thus is proportional to the product of the vertical speed
passing through the plane, $|W_{z=0}|$, and the vertical oscillation
period for that star's orbit.  The vertical period should be
calculated from the measured vertical potential gradient of the
Galactic disk rather than from an harmonic potential approximation
(which has constant vertical period).  Figure 9 shows the relation
betweeen oscillation period and vertical speed that results from the
vertical force law derived by Kuijken and Gilmore (1989). This was
adopted here.

In principle the correction could be done on a star-by-star basis,
using the space motions in the Gliese catalog.  However, the small
sample size, 91 stars after pruning, leaves this procedure susceptible
to sampling errors.  A more robust weighting is that derived from the
best available estimator of the true vertical velocity dispersion
appropriate for a specific abundance range, provided that one knows
reliably the relative fractions of stars in that specific abundance
range which are thin disk and thick disk. As emphasised earlier, this
information is not available {\it a priori}. Thus weightings of this type are
inherently unreliable. However, to compare with previous
analyses we go through such a weighting exercise here.

The vertical velocity dispersion as a function of iron abundance
for the Gliese catalog sample -- both the `raw' sample and that
obtained subsequent to pruning of potentially short-lived stars --
together with that of the local F/G star sample of Edvardsson \etal\
(1993) is shown in Figure 10.  The solid curve is the result of
convolving a step-function relationship between velocity dispersion
and [Fe/H], with the transition between a velocity dispersion of 19 km/s to
42 km/s at $-0.4$ dex, as suggested by Figure 7, with a
Gaussian of dispersion 0.1 dex, the expected metallicity accuracy.
The smooth dashed curve results from adopting a value of $-0.5$ dex
for the transition metallicity, and serves as an estimate of the
uncertainty introduced by this approach.  These smooth relationships
wer used to assign a velocity dispersion to each metallicity bin of
the present sample, rebinned into 0.2 dex wide bins.  The kinematic
correction were thus effected by weighting the binned data (after
evolutionary corrections) by the product of the vertical velocity
dispersion of that bin, times the extra factor due to the period
lengthening in a non-harmonic potential (from Figure 9), adopting the
velocity dispersion as the estimator for the relevant mean vertical
speed; this is justified for {\it relative\/}
weighting.\footnote{$^\heartsuit$}{Note also that using our procedure
with the adopted velocity dispersions of Sommer-Larsen (1991) resulted
in a weighting factor that is numerically equal (to within a few
percent) to that obtained by Sommer-Larsen by his more indirect
technique.}  The resulting metallicity distribution was then
re-normalised to the same number of stars, for convenience.  This
procedure resulted in the metallicity distribution of Figure 11, where
the solid and dashed histograms correspond to the adopted kinematics
in the weighting scheme indicated by the solid and dashed curves in
Figure 10.  One should remember that the kinematic weighting assumes
that all stars more metal-poor than $-0.4$ dex (or $-0.5$ dex) belong
to a single population, with kinematics characterized by a simple,
single Gaussian; this is a dubious assumption.

The metallicity distributions of
Figure 11 do  not differ radically from earlier determinations.
There are  somewhat fewer stars below [Fe/H] $\sim -0.7$ dex than
have the distributions resulting from Cameron's (1985) UV excess
calibration from broadband UBV photometry (Pagel 1989 and
Sommer-Larsen 1991); as we have noted this metallicity calibration
provides systematically lower [Fe/H] estimates than \eg\ that of
Carney (1979), while the larger errors of the UV excess technique will
broaden the distribution.  After weighting, the present sample has
relatively fewer stars above solar metallicity than does either of the
Pagel distributions, but is similar to that of Sommer-Larsen.  The
fraction of stars more metal-poor than $-0.4$ dex is approximately 45\%.

For interest, to compare with previous results, we show in Figure
12(a) these same distributions, with an overlaid Simple Model of yield
0.7 solar abundance, and evolved to a gas fraction of 0.25, as
appropriate for the local disk (Kuijken and Gilmore 1989).  While the
model could not be described as providing an excellent dscription of
the data, the discrepancies are not overwhelming.  As we emphasise
below, however, we explicitly do not suggest that this model is a
viable representation of these data, nor that the distribution
function of Figure 12(a) is that appropriate to the local Galactic
disk.  Indeed, the possible traps
for the unwary may be illustrated by, following Pagel (1989),
transforming the iron abundance distribution into one for oxygen. The
Simple Model predictions are based on the instantaneous recycling
approximation, and as such are better compared to metallicity
distributions based on elements that are produced predominantly by
massive stars, such as oxygen, rather than the iron-based distribution
usually measured (cf. Pagel 1989).  As discussed above, for most of
the stars of interest, with [Fe/H] $\simgt -1$ dex, the simple
relationship [O/H]$ = -0.5$[Fe/H] is a reasonable approximation
(e.g. Edvardsson {\it et al.}  1993).  Using this to transform an iron
distribution into an oxygen distribution simply squashes the
metallicity scale of the observations.
The resulting distribution and Simple
Model are shown in Figure 12b; it is much clearer now that
the Simple Model distribution is significantly {\it wider\/} than the
observations.  \footnote{$\diamondsuit$}{It should be noted that the gas
fraction appropriate to
the solar circle, 0.25, is rather higher than had been adopted in the
past; this larger value results in the truncation of the model
evolution at metallicities below the extent of the data, for yields
that approximate the location of the peak in the observational data.
The main input parameters
to the local gas fraction whose values have been re-estimated recently
are the actual gas surface densities, and the total disk surface
density (see Kuijken and Gilmore 1989). }

\bigskip
\centerline {\it 3.2 Comparison to  the {\it in situ} Sample}

Gilmore, Wyse and Jones (1995) determined the iron abundance
distribution of G-dwarfs at several kpc above the Galactic plane.
Those distributions were derived from spectroscopic observations of
area-complete, photometrically-defined samples of F/G dwarfs in the
apparent magnitude range $16 \simlt V \simlt 18$.  The final
distribution of stars {\it per unit volume\/} at a chosen fiducial
distance depends on the density law assumed for the stars.  Figure 19
from that paper is reproduced here as Figure 13, showing the iron
abundance distributions at $z = 1.5$kpc in the direction of the South
Galactic Pole and at $z = 1$kpc in a field at $(\ell = 270,\, b =
-45)$, UK Schmidt F117; for the magnitude range under study these two
lines-of-sight probe essentially the same Galactocentric distance
projected on the plane.  The distributions shown are derived under two
sets of assumed density laws; in one case that all the stars are thick
disk stars, and in the other case that the stars more metal-poor than
${\rm [Fe/H]=-1.0}$ are halo objects, while the stars more metal-rich
than ${\rm [Fe/H] = -0.4}$ are 90\% thick disk and 10\% thin disk
stars (a ratio consistent with star-count model predictions at the
sample distance from the Galactic plane) and 100\% thin disk for ${\rm
[Fe/H] > 0.0}$.  Adopted vertical scale heights were 4000pc for the
halo, 1000pc for the thick disk, and 250pc for the thin disk, in
agreement with the adopted velocity dispersions and vertical potential
above. Full details are provided in Gilmore, Wyse and Jones (1995).

The metal-rich bins are expected to be under-represented, since the
sample was selected {\it in situ\/}, far from the Galactic plane,
requiring a high minimum vertical velocity at the plane, for a given
star. Metal-rich stars in general may be expected to be drawn from a
population with vertical velocity dispersion of only $\sim 20
\kms$ (Edvardsson \etal\  1993; Freeman 1991), so that only stars in
the tail of the velocity distribution will be sampled {\it in situ\/}
at a few kpc.  That is, this sample suffers from the complementary
selection bias to that suffered by the nearby-star sample,
systematically undersampling low velocity-dispersion populations.
Indeed, there is {\it no\/} significant dependence of the observed
line-of-sight velocity dispersion on metallicity for either the SGP
stars (typical dispersion is $\sigma_W \sim 40 \kms$), or the F117
stars (typical dispersion is $\sigma_{los} \sim 50 \kms$; Gilmore and
Wyse, { in preparation}), suggesting that the colder thin-disk stars
are indeed relatively under-sampled.  We can use this kinematic bias to
determine the number of low vertical velocity, metal-poor stars in the
local sample.

The predicted distribution at $z=0$pc may be derived from these data,
again once an appropriate choice of density law is made.  Obviously
the case where all stars are thick-disk stars, and so each metallicity
bin is corrected using the same density profile, will have the same
relative distribution as that shown in Figure 13, with only the
overall normalisation increased. The distributions obtained under the
alternative assumption, that the metal-poor bins are dominated by halo
stars, while the metal-rich bins are dominated by thin-disk stars,
are shown in Figure 14.  It should be noted that the original sample
has very few stars at either extreme in metallicity.

The two lines-of-sight provide
independent probes of the {\it in situ\/}  metallicity distribution,
with approximately equal numbers of stars in the original samples, and
in each bin, and so a composite distribution may be obtained by a
simple unweighted sum.

The distributions shown are most reliable in the range of $-1 \simlt
{\rm [Fe/H]} \simlt -0.2$, where we have most stars in the original
sample, and for which the iron-abundance estimator is most robust (see
Jones, Gilmore and Wyse 1995; Jones, Wyse and Gilmore 1995; Gilmore,
Wyse and Jones 1995 for details).  This is also the range of
metallicity for which there is minimal difference between the two
treatments of the density law(s).  Comparing to the unweighted Gliese
sample of Figure 5, the {\it in situ\/} sample at the plane has a
rather different distribution for metal-poor stars, being flat from [Fe/H]$
\sim -0.5$ to [Fe/H] $ \sim -1$ dex.  The difference in distributions
between the local sample and the {\it in situ\/} sample brought down
to the plane is due to the presence of those kinematically-cold stars
in the local sample with metallicity [Fe/H]$ \simlt -0.4$.  Thus
comparison of these two samples thus allows us to deduce the shape of
the thin disk metallicity distribution below its peak.

The distributions functions must first be matched: this should be at
an iron abundance where the thick disk may reasonably be expected to
dominate, rather than just be present as a small minority population.
An independent estimate of this abundance may be made by inspection of
the age-metallicity-kinematics relations of Edvardsson {\it et al}
(1993) and of Marquez and Schuster (1994).  These show that, although
there is a substantial scatter in the thin disk age-abundance
relationship, there is no significant population of stars with ${\rm
[Fe/H]} \simlt -0.7$ dex that is younger than 12Gyr, or has kinematics
cooler than those of the thick disk.  The stellar halo, on the other
hand, dominates for ${\rm [Fe/H]} \simlt -1$ dex (Marquez and Schuster
1994).  Thus, we match the {\it in situ\/} sample, scaled to the
plane, and the local sample, over the interval $-0.7 > {\rm [Fe/H]} >
-1$ dex.  Further, both distributions should be reliable here. We
carry out this matching for both scalings to the plane, that assuming
that all stars at $z=1500$pc are thick disk stars, and that
assuming  the
three-component distribution in the {\it in situ \/} sample.
The uncertainties in the iron abundance estimates of the {\it
in situ\/} sample are around 0.2 dex; the local sample was
re-binned to this prior to the matching.

In view of the small number statistics in the local sample at low
abundances, we proceed by adopting the {\sl shape} of the thick disk
chemical abundance distribution function as derived from the scaled
{\it in situ} sample. Given that {\sl shape}, we normalise the counts
iteratively to provide a self-consistent match to the local metal-poor
abundance distribution and the kinematic distributions shown earlier.
That is, the normalisation of the thick disk distribution function is
derived by allocating a subset of those stars in the abundance range
$-0.8 < $[Fe/H] $< -0.4$ to the thick disk, with the {\sl relative}
numbers from each abundance bin determined by the adopted shape for
the thick disk distribution. The normalisation is derived from the
combination of the kinematics of Figure 9 and the (small) number of
stars in the chemical abundance range $-1.0 <$ [Fe/H] $< -0.8$, which
is assumed pure thick disk.

This matching results in the histograms shown in Figure 15. The dashed
lines denote the $z = 0$ pc, formerly {\it in situ,} distribution,
assuming all stars are thick-disk stars; the dash-dotted curve is the
comparable distribution adopting the three-component scaling,
as
defined above -- both
now re-normalised as above to match
the solid histogram, which represents the local Gliese
long-lived F/G dwarf sample. While the normalisation is imprecise in
consequence of small-number statistics, it is clear that, with either
assumption in re-normalising the {\it in situ \/} sample, there is a
considerable excess group of stars with [Fe/H]$\sim -0.5$ in the local
Gliese sample, compared to the number in the other distributions. It is
also apparent that there is reasonable agreement between the
re-normalised {\it in situ \/} sample, after scaling to the plane, in
the number of stars with $-0.4 \simlt {\rm [Fe/H]}
\simlt 0.0$, although the detailed shape of the distribution function
remains uncertain, {\sl only } if the three-kinematic component model
is adopted. We note that it is this model which is also consistent
with star count models (Gilmore, Wyse and Jones 1995).

That is, a matching of the local volume-complete sample of long-lived
F/G dwarfs with a comparable sample, selected near 1500pc from the
Sun, and corrected through density profiles and normalizations
consistent with kinematic and star count data, requires that a
majority of the stars near the Sun with [Fe/H] near $-0.5$dex have
kinematics appropriate to the thin disk and not the thick disk. This
population  may  extend, but only  as a minority constituent,
to at least
[Fe/H] $ \approx -0.8$. Similarly, it is probable that some stars with
thick disk kinematics extend upwards in metallicity to [Fe/H] $\approx
-0.2$dex.  It is this overlap of populations which resolves the
apparent inconsistency between star count models, which require the
local thick-disk normalization in a volume--complete sample to be a few
percent, and the abundance distribution of the local volume--complete
(Gliese catalog) samples, where metal-poor stars comprise some 25
percent of the total.

In summary, from this deconvolution, we now derive an abundance
distribution for the long-lived stars of the Galactic disk by
assigning the stars with ${\rm -0.8 < [Fe/H] \leq -0.6}$ to
populations with vertical velocity dispersions of 42km/s and 19km/s
(Freeman 1991) in the ratio of 3:2; those in the interval ${\rm -0.6 <
[Fe/H] \leq -0.4}$ in the ratio 1:5; and those in the interval ${\rm
-0.4 < [Fe/H] \leq -0.2}$ in the ratio 1:10. With these kinematic
assignments, we are able to derive a column-integral abundance
distribution, utilising the weightings of Figure 9 above.

For completeness, we also recall the results of Wyse and Gilmore
(1992) and of Ibata and Gilmore (1995), which show that the angular
momentum distributions of the halo and the inner bulge are
indistinguishable from each other, but are very different from those
of the thick disk or the thin disk, which are in turn
indistinguishable from each other.  Thus, those studies conclude that
the halo -- bulge system of the Galaxy is a parallel evolutionary
sequence to that of the thick disk -- thin disk. We summarise all
these results in Figure 16. This shows the (normalised to unity for
each distribution) abundance distribution of
the Solar neighborhood halo (Laird \etal 1988), the outer
Galactic bulge (Ibata and Gilmore 1995), the younger stars of the
solar neighborhood, from Figure 6 above, the volume complete sample of
long-lived thin disk stars, derived here from our composite Gliese and
scaled {\it in situ\/} samples, the (similarly derived) volume
complete sample of local thick disk stars, and the column-integral
summed abundance distribution for the long-lived thin disk and the thick
disk derived here from those latter two distributions. The six
distributions are also presented, for convenience, in Table 3.

\bigskip
\centerline{ 4. DISCUSSION}

A variety of distribution functions relevant to the evolution of the
Galactic disk, and in particular to the question of the stellar
population structure in the Galactic disks, have become reliably
defined in the last few years. Among the most important of these have
been determinations of the age-metallicity-kinematics distribution of
F/G stars in the Solar neighborhood. These have been complemented by
many recent star-count and kinematic studies, which have determined
reasonably reliable and precise normalisations for the three stellar
populations evident in the Solar neighborhood data: the halo, the
thick disk and the thin disk. In this paper we have achieved a
new derivation of
the
abundance distribution for the local, volume-complete sample of
long-lived stars, using the most recent and precise photometry and
calibrations. This determination, which is in general agreement with other
recent such determinations, but more reliable, was
demonstrated above to provide a local fraction of metal-poor
stars which exceeds that derived from the larger-scale studies
(star counts and {\it in situ\/} chemical abundance
distributions) by a factor of about five. Comparison of the local
abundance determination with that determined from two {\it in
situ\/} samples, each some 1500pc from the Sun, isolates the
explanation of this inconsistency: a majority of the stars with
metallicity near [Fe/H]$\approx -0.5$dex in the Solar
neighborhood have thin disk kinematics. The apparent rather
abrupt transition in  kinematics
near [Fe/H]$\approx -0.4$dex, as in  Figure 7,  is rather a confusing
manifestation of two overlapping distribution functions.  The
combination of kinematics, spatial distribution and chemical
abundances for samples with different selection effects was
required for this understanding.

The underlying  metallicity distribution function for
the old thin disk extends to at least [Fe/H]$\approx -0.75$dex, and
possibly to even lower abundances. Similarly, the thick disk abundance
distribution extends up to at least [Fe/H]$\approx -0.4$dex, and
possibly to even higher abundances. This overlap, while confusing
simple analyses, is consistent with the observed scatter in recent
age-metallicity relations, which show a significant number of stars
with [Fe/H]$\simlt -0.5$dex, and with ages of only a few Gyr, and
which are providing improving evidence for stars with [Fe/H]$\approx
0$, and with ages more typical of the bulk of the thick disk, some
12Gyr (cf. Figure 14(a) of Edvardsson {\it et al.} 1993).  It is clear
that age distributions are are also going to provide important
constraints on theories of disk evolution.
Whether these are old, thin disk stars or metal-rich thick
disk stars remains a matter for future investigations.

The result that  the abundance distributions of the thick
disk, which is apparently exclusively old, and the thin disk overlap
considerably, with the thin disk extending to abundances nearly as
metal-poor as does the thick disk has implications for models of the
formation of the thick disk. This is consistent with the
simplest class of merger models for the formation of the thick disk
only if there is substantial scatter in the age-metallicity
relationship for thin disk stars at all times, even during the
earliest stages. In such models the thick disk is formed during the
merger of a predominantly stellar satellite galaxy with an extant
Galactic disk (e.g. Quinn, Hernquist and Fullagher 1993). Dramatic
kinematic heating of those thin-disk stars already formed abruptly
thickens the thin disk into a thick disk, which will also contain much
of the stellar population of the incoming merging galaxy. The
remaining disk gas will rapidly re-stabilise into a new thin disk, and
continue its evolution. The chemical evolution is relatively
unaffected by the merger, and the first newly-forming thin-disk stars
will have the mean metallicity which the interstellar medium had
attained prior to the merger, which will be greater than the mean
abundance of the stars in the newly-thickened disk. In this class of
model, substantial overlap in abundances is not a natural feature
without appeal to large scatter in the age-metallicity relationship of
those stars formed prior to the merger, as well as those formed
after.

As part of the present analysis, we have isolated, in order to remove
from our complete sample of long-lived stars, the distribution of
abundances of younger stars near the Sun. This distribution is more
metal-rich in the mean, and is skewed to higher abundances, than is
the distribution of abundances of longer-lived stars, which will in
the mean be older.  Thus, there is a trend of abundance with age in
the Galactic disk. Remarkably, however, there appears to have been
little or no change in the mean abundance of the old disk in the
interval between about 10Gyr ago and about 3Gyr ago. In the last few
Gyr the mean iron abundance has increased by a factor of about two,
while the tail of the thin disk abundance distribution to metal-poor
stars has disappeared.  This conclusion from the present analysis is
consistent with the age-metallicity-kinematic data of Edvardsson {\it
et al.} (1993), which show a turn-up in the mean chemical abundance
within the last few Gyr (their Figure 14(a)).

Identification of a large fraction of the metal-poor stars in the
Solar neighborhood with the thin disk, so that their kinematics
confine them close to the disk plane, means that recent attempts
(including those by ourselves in section 3.1 above) to derive the
column-integral abundance distribution near the Sun by
kinematic-weighting of the local data will {\sl over-estimate} the
true number of metal-poor stars in the Galactic disk.  This arises
since the practice has been to identify all metal-poor stars with the
thick disk, and thus necessarily use thick disk kinematics in the
weighting applied to all these stars. Figure 8  demonstrated the danger
of assuming a simple  single
Gaussian fit to determine  the kinematics
appropriate to a given sample.  Hence, to use the jargon of the Simple
Model of chemical evolution, the `G-dwarf problem' seen in the fitting
of the solar cylinder data is even more severe than has been suggested
recently, since the true contribution of metal-poor stars to the
column integral has been over-estimated in the past by the simple
kinematic weighting procedure which had been adopted. We
emphasise, however, that the assumptions underlying the Simple
Model are inappropriate to the evolution of a part of a galactic
disk, so that this deduction by itself is not a very helpful clue
to improve models of galactic evolution.

A variety of recent studies have provided a determination of the Solar
cylinder chemical abundance distribution taking account of local
kinematics. In this study we have included data for distant stellar
samples, allowing a more robust deconvolution of the high-precision
but small sample local data than has been possible heretofore. Such
abundance determinations are useful constraints on Galactic chemical
evolution, but contain insufficient information for detailed
analysis. Further progress requires consideration of the multi-variate
distribution function, complementing the results of this study with
both age and chemical element ratio information.

\bigskip

\centerline {ACKNOWLEDGEMENTS}

 We thank the referee, Bernard Pagel, for helpful suggestions which
have improved the presentation.  RFGW acknowledges support from the
AAS Small Research Grants Program in the form of a grant from NASA
administered by the AAS, from the NSF (AST-9016226) and from the
Seaver Foundation.  Our collaboration was aided by NATO Scientific
Affairs in the very early stages and currently by the NSF
(INT-9113306).  NSSDC provided necessary access to data catalogs.  The
Center for Particle Astrophysics is supported by the NSF.
\bigskip
\centerline {REFERENCES}
\parindent=0pt

\pp Aguilar, L., Carney, B., Latham, D. and Laird, J. 1995, in preparation

\apjref Bergh, S. van den 1962;\aj;67;486

\apjref Cameron, L.M. 1985;\aap;146;59
\apjref Carney, B. W., 1979;\apj;233;211
\apjref Carney, B. W., Latham, D.L. and Laird, J. 1989;\aj;97;423
\apjref Crawford, D.L. and Barnes, J.V. 1969;\aj;74;407
\apjref Edmunds, M.G. 1990;\mnras;246;678
\apjref Edvardsson, B., Andersen, J., Gustafsson, B., Lambert, D.L., Nissen,
P. and Tomkin, J. 1993;\aap;275;101

\pp Freeman, K.C. 1991, in `Dynamics of Disc Galaxies', ed B.~Sundelius
(G\"oteborgs University, G\"oteborg) p15.

\pp Freeman, K.C. 1994, in `Dynamics of Disc Galaxies', special
session at the AAS Summer meeting 1994.

\pp Gilmore, G. 1990, `The Milky Way as a Galaxy', (University
Science, Berkeley), G.~Gilmore, I.~King, and P.C.~van der Kruit.

\apjref Gilmore, G. and Wyse, R.F.G. 1985;\aj;90;2015
\apjref Gilmore, G. and Wyse, R.F.G. 1986;Nature;322;806

\apjref Gilmore, G., Edvardsson, B. and Nissen, P. 1991;\apj;378;17

\pp  Gilmore, G., Wyse, R.F.G. and Jones, J.B. 1995, AJ (in press)

\apjref Gilmore, G., Wyse, R.F.G. and Kuijken, K. 1989;\araa;27;555

\apjref Hartwick, F.D.A. 1976;\apj;209;418

\pp Ibata, R.A., and Gilmore, G., 1995. \mnras, in press.

\pp Jones, J.B., Gilmore, G. and Wyse, R.F.G. 1995, \mnras, submitted

\pp Jones, J.B., Wyse, R.F.G. and Gilmore, G., 1995, \pasp, submitted

\apjref Kuijken, K. and Gilmore, G. 1989;\mnras;239;605

\pp Lacey, C.G. 1991, in `Dynamics of Disc Galaxies', ed B.~Sundelius
(G\"oteborgs University, G\"oteborg) p257

\apjref Laird, J., Rupen, M.P., Carney, B. and Latham, D. 1988;\aj;96;1908

\apjref Magain, P., 1987;\aap;179;176

\apjref Majewski, S.R. 1993;\araa;31;575

\apjref Marquez, A. and Schuster, W.J., 1994;\aaps;108;341

\apjref McWilliam, A. 1990;\apjs;74;1075
\apjref Nissen, P.E., Gustaffson, B., Edvardsson, B. and
Gilmore, G. 1994;\aap;285;440

\apjref Norris, J. and Ryan, S. 1989;\apj;340;739
\apjref Norris, J. and Ryan, S. 1991;\apj;380;403

\apjref Ojha, D.K., Bienayme, O., Robin, A.C. and Mohan, V. 1994;\aap;290;771
\apjref Olsen, E.H. 1983;\aaps;54;55
\pp Pagel, B.E.J. 1989, in `Evolutionary Phenomena in Galaxies',
eds J.E.~Beckman and B.E.J.~Pagel (CUP: Cambridge) p201.

\apjref Pagel, B.E.J. and Patchett, B.E. 1975;\mnras;172;13

\apjref Quinn, P.J., Hernquist, L. and Fullagher, D.P., 1993;\apj;403;74
\pp Ryan, S.G. and Norris, J.E.  1993, in `Galaxy Evolution: The Milky
Way Perspective' ed S.~R.~Majewski (ASP, San Francisco) p103.

\apjref Sandage, A. 1969;\apj;158;1115
\apjref Saxner, M. and Hammarb\"{a}ck, G., 1985;\aap;151;372

\apjref Schmidt, M. 1963;\apj;137;758
\apjref Schuster, W.J. and Nissen, P.E. 1989a;\aap;221;65
\apjref Schuster, W.J. and Nissen, P.E. 1989b;\aap;222;69

\apjref Sommer-Larsen, J. 1991;\mnras;249;368
\apjref Sommer-Larsen, J.  and Antonuccio-Delugo, V. 1993;\mnras;262;350

\apjref Tinsley, B. 1975;\apj;197;159
\apjref Tosi, M. 1988;\aap;197;33
\apjref Truran, J.W. and Cameron, A.G.W., 1971;Ap Sp Sci;14;179

\apjref VandenBerg, D.A. 1985;\apjs;58;711
\apjref VandenBerg, D.A. and Bell, R.A. 1985;\apjs;58;561
\apjref VandenBerg, D.A.  and Laskarides, P.G. 1987;\apjs;64;103

\pp Wyse, R.F.G. 1995, in `Stellar Populations', IAU Symposium 164,
eds P.C.~van der Kruit and G.~Gilmore (Kluwer, Dordrecht), in press

\apjref Wyse, R.F.G. and Gilmore, G. 1992;\aj;104;144
\vfill\eject
\header {FIGURE CAPTIONS}

\pp
Figure 1 (a) : The solar cylinder metallicity distribution from Pagel
and Patchett (1975).

\pp
Figure 1 (b) : The same data as (a), but with a different metallicity
calibration, from Pagel (1989).

\pp
Figure 1 (c) : The same data and metallicity calibration as (b), but with a
different kinematic weighting, from Sommer-Larsen (1991).

\pp
Figure 2 : Metallicity estimated using the technique of Edvardsson \etal\
against that obtained using the technique of Schuster and Nissen.

\pp
Figure 3 : Metallicity distribution of all local F/G dwarfs selected from the
cross correlation of the Gliese and Olsen catalogs, plus the Sun.

\pp
Figure 4 (a) : The sample of Figure 3 in the $c_1 - (b-y)$ plane,
together with turnoff points for a range of metallicities, as described in
the text.

\pp
Figure 4 (b) : The local sample after removal of stars expected to
have main-sequence lifetimes shorter than the age of the disk.

\pp
Figure 5 : Metallicity distribution of the local sample after the
removal of short-lived stars.

\pp Figure 6 : Histograms of the iron abundance distributions for
the Edvardsson \etal\ sample with ages less than 12 Gyr, after
corrections for completeness (dashed lines); the McWilliam local G/K
giant sample (dot-dashed lines) and the present sample of long-lived
F/G dwarfs  from
the Gliese catalog (solid lines).

\pp Figure 7 : [Fe/H] rank versus absolute value of the vertical
velocity for the Edvardsson \etal\ sample, following Freeman (1991).
The straight dashed line has been drawn to illustrate the change in
kinematics.

\pp Figure 8 (a) : Histogram of the W-velocity data for the
subsample of the Edvardsson {\it et al.} stars with iron abundance in
the range $-0.4 \ge {\rm [Fe/H]} > -0.9$, where naively one might have
expected the thick disk to dominate.  The smooth curve is a Gaussian
of dispersion 45 km/s, normalised by eye.

\pp Figure 8 (b) : As (a), but for the complete sample of
Gliese F/G dwarfs.

\pp Figure 8 (c) : As (a), but for the more metal-poor stars,
with iron abundance in
the range $-0.5 \ge {\rm [Fe/H]} > -0.9$.

\pp Figure 8 (d) : As (b), but for the more metal-poor stars,
with iron abundance in
the range $-0.5 \ge {\rm [Fe/H]} > -0.9$.

\pp
Figure 9 : Ratio of the  vertical orbital period in the actual
Galactic potential to that in an harmonic potential, as a function of  vertical
velocity at the disk plane.

\pp
Figure 10 : Vertical velocity dispersion as a function of iron
abundance for the Gliese catalog stars and the F/G star sample of
Edvardsson \etal\ (1993). The star symbols refer to the `raw' Gliese
sample, the open squares to that obtained subsequent to pruning of
potentially short-lived stars, and the filled circles to the
Edvardsson \etal\ sample. The error bars are appropriate for an
underlying Gaussian distribution.  The solid line is the result of
convolving a Gaussian of dispersion 0.1 dex with the step-function
velocity-dispersion -- metallicity relation suggested by the data.

\pp
Figure 11 :  Iron abundance   distribution of the local sample after the
weighting to account for kinematic bias.

\pp
Figure 12(a) : As Figure 11 but with the predictions of a `simple closed-box'
model of yield 0.7Z$_\odot$, and present gas fraction 0.25, truncated
at the metallicity corresponding to these values and normalised to the
total number of stars in the sample.

\pp Figure 12(b) : As (a), but for the transformed `oxygen'
metallicity distribution.

\pp Figure 13 : Figure 19 from Gilmore, Wyse and Jones (1995).
Comparison of the F117 $z=1000$pc iron abundance distribution (solid
histograms) and the SGP $z=1500$pc distribution (dashed histograms).
The left panel shows the derived distributions under the assumption
that all stars follow the thick disk density profile, while the right
panel includes thin disk and stellar halo contributions.

\pp Figure 14 : The $z=0$pc iron abundance distribution resulting from
the {\it in situ\/} data, if density laws appropriate to the halo are
assumed for the metal-poor stars, and that appropriate to the thin disk for the
most metal-rich stars, as described in the text.

\pp Figure 15 : The solar neighborhood  iron-abundance distribution that
results from matching the {\it in situ\/} data, brought to the Plane,
to the local data, as described in the text.  The dashed lines denote
the $z = 0$ pc, formerly {\it in situ,} distribution, assuming all
stars are thick-disk stars; the dash-dotted curve is the comparable
distribution adopting the three-component scaling -- both now
re-normalised to match, over the interval $-0.8 > {\rm [Fe/H]} > -1$
dex, the solid histogram, which represents the local Gliese long-lived
F/G dwarf sample. The {\it in situ\/} data have been truncated at solar
metallicity.

\pp Figure 16 : Abundance distributions, normalised to unity, of, from
top to bottom,
the Solar neighborhood halo (Laird \etal\ 1988); the outer
Galactic bulge (Ibata and Gilmore 1995), truncated by them at solar
metallicity; the younger stars of the
solar neighborhood, from Figure 6 above; the volume complete sample of
long-lived thin disk stars, derived here from our composite Gliese and
scaled {\it in situ\/} samples; the (similarly derived) volume
complete sample of local thick disk stars; and the column-integral
 abundance distribution for the sum of
the long-lived thin disk and the thick
disk, derived here from those latter two distributions.

\vfill\eject
\parskip=3pt plus 1pt minus 1pt  
  \tabskip=.2em plus .5em minus .2em
  
  $$ \vbox{ \halign to \hsize { \hfil#\hfil & \hfil#\hfil & \hfil#\hfil &
  \hfil#\hfil & \hfil#\hfil & \hfil#\hfil & \hfil# & \hfil#\hfil &
\hfil# \cr
 \multispan{9} \hfil {Table 1 : {Gliese Catalog F/G Dwarfs}}
\hfil \cr
 \noalign{ \vskip 12pt } \noalign{ \hrule height 1pt \vskip 1pt \hrule
 height 1pt } \noalign{ \vskip 8pt } HD & B$-$V & $(b-y)$ & $ m_1 $ &
 $c_1$ & $\beta$ & [Me/H] & rejected? & W vel \cr
\noalign{
 \vskip 8pt\hrule height 1pt \vskip 8pt }
1273 & 0.64 & 0.409 & 0.182  & 0.248 & 2.592 & $-$0.57 & no & $-$5 \cr
1388 & 0.59 & 0.379 & 0.191 & 0.345 & \dots & $-$0.01 & yes & \dots \cr
3079 & 0.54 & 0.364 & 0.156 & 0.367 & 2.607 & $-$0.41 & no & 12 \cr
3823 & 0.56 & 0.359 & 0.156 & 0.348 & 2.609 & $-$0.38 & yes & \dots \cr
4208 & 0.67 & 0.404 & 0.227 & 0.278 & \dots & $-$0.10 & no & \dots \cr
6434 & 0.60 & 0.383 & 0.159 & 0.275 & 2.586 & $-$0.53 & no & $-$10 \cr
10800 & 0.61 & 0.390 & 0.201 & 0.302 & \dots & $-$0.10 & yes & \dots \cr
11112 & 0.65 & 0.409 & 0.200 & 0.400 & \dots & $-$0.05 & no & \dots \cr
13043 & 0.61 & 0.388 & 0.206 & 0.379 & \dots & \phantom{$-$}0.14 & yes \cr
13445 & 0.82 & 0.486 & 0.338 & 0.283 & \dots & $-$0.17 & no & \dots \cr
14412 & 0.73 & 0.445 & 0.247 & 0.231 & \dots & $-$0.52 & no & $-$10 \cr
14802 & 0.60 & 0.390 & 0.185 & 0.379 & \dots & $-$0.11 & yes & \dots \cr
16141 & 0.67 & 0.424 & 0.211 & 0.377 & 2.594 & $-$0.05 & no & \dots \cr
16397 & 0.60 & 0.388 & 0.156 & 0.281 & 2.578 & $-$0.58 & no & $-$31  \cr
16417 & 0.65 & 0.413 & 0.205 & 0.397 & \dots & $-$0.02 & no & \dots \cr
16784 & 0.56 & 0.380 & 0.131 & 0.302 & 2.597 & $-$0.82 & no & 22 \cr
17051 & 0.56 & 0.356 & 0.190 & 0.363 & 2.637 & \phantom{$-$}0.10 & yes \cr
17865 & 0.55 & 0.371 & 0.141 & 0.309 & 2.593 & $-$0.61 & no & $-$23  \cr
17925 & 0.87 & 0.511 & 0.390 & 0.297 & \dots & $-$0.14 & no & \dots \cr
18803 & 0.72 & 0.429 & 0.267 & 0.326 & \dots & \phantom{$-$}0.16 & no & \dots
\cr
19467 & 0.65 & 0.409 & 0.200 & 0.337 & \dots & $-$0.13 & no & \dots \cr
20165 & 0.86 & 0.508 & 0.407 & 0.297 & \dots & $-$0.14 & no & \dots \cr
\noalign{ \vskip 8pt\hrule height 1pt \vskip
 8pt } }}$$

   \vfill\eject
  $$ \vbox{ \halign to \hsize { \hfil#\hfil & \hfil#\hfil & \hfil#\hfil &
  \hfil#\hfil & \hfil#\hfil & \hfil#\hfil & \hfil# & \hfil#\hfil &
\hfil# \cr
 \multispan{9} \hfil {Table 1 (continued) }
\hfil \cr
 \noalign{ \vskip 12pt } \noalign{  \hrule
 height 1pt } \noalign{ \vskip 8pt } HD & B$-$V & $(b-y)$ & $ m_1 $ &
 $c_1$ & $\beta$ & [Me/H] & rejected? & W vel  \cr
 \noalign{ \vskip 12pt }
 \noalign{ \hrule height 1pt }
 \noalign{ \vskip 8pt }
   20407 & 0.58 & 0.373 & 0.162 & 0.280 & 2.605 & $-$0.40 & no & $-11$ \cr
   20619 & 0.66  & 0.405   & 0.215   & 0.276   & \dots   & $-$0.19  & no &
\dots \cr
   20794 & 0.71  & 0.440   & 0.235   & 0.283    & 2.575   & $-$0.26  & no &
\dots \cr
   22049 & 0.88  & 0.514   & 0.410   & 0.273   & \dots   & $-$0.24  & no &
\dots \cr
   22484 & 0.57  & 0.370   & 0.169   & 0.379   & \dots   & $-$0.30  & yes &
\dots \cr
   22879 & 0.54  & 0.369   & 0.120   & 0.273    & 2.580   & $-$0.81  & no &
$-$49 \cr
   24293 & 0.66  & 0.412   & 0.211   & 0.332   & \dots   & $-$0.07  & no &
\dots \cr
   30495 & 0.63  & 0.399   & 0.213   & 0.320    & 2.593   & \phantom{$-$}0.00 &
yes & \dots \cr
   36435 & 0.77  & 0.454   & 0.300   & 0.263   & \dots   & $-$0.20 &
no & \dots  \cr
   36767 & 0.54  & 0.344   & 0.168   & 0.336   & \dots   & $-$0.14   & yes &
\dots \cr
   37655 & 0.60  & 0.380   & 0.176   & 0.373    & 2.593   & $-$0.17   & yes &
\dots \cr
   42618 & 0.63  & 0.402   & 0.219   & 0.303    & 2.581   & $-$0.03   & yes &
\dots \cr
   42807 & 0.67  & 0.414   & 0.228   & 0.293    & 2.582   & $-$0.09 &
no & \dots  \cr
   43834 & 0.72  & 0.442   & 0.263   & 0.339    & 2.601   &
\phantom{$-$}0.11 & no & \dots   \cr
   46588 & 0.50  & 0.345   & 0.158   & 0.346   & \dots   & $-$0.27  &
yes & \dots  \cr
   50806 & 0.72  & 0.432   & 0.235   & 0.368   & \dots   &
\phantom{$-$}0.08 & no & \dots  \cr
   52449 & 0.53  & 0.332   & 0.177   & 0.391   & \dots   & \phantom{$-$}0.06
& yes & \dots \cr
   52711 & 0.60  & 0.386   & 0.182   & 0.330    & 2.590   & $-$0.18   & no &
\dots \cr
   56274 & 0.61  & 0.373   & 0.181   & 0.246    & 2.575   & $-$0.17  &
yes & \dots  \cr
   61994 & 0.66  & 0.445   & 0.247   & 0.298   & \dots   & $-$0.16  & no &
\dots \cr
   62613 & 0.73  & 0.450   & 0.261   & 0.293    & 2.567   & $-$0.14   & no &
\dots \cr
   62644 & 0.77  & 0.493   & 0.226   & 0.382   & \dots   & $-$0.36   & no &
\dots \cr
    \noalign{ \vskip 8pt\hrule height 1pt
  \vskip 8pt }
}}$$

   \vfill\eject

  $$ \vbox{ \halign to \hsize { \hfil#\hfil & \hfil#\hfil & \hfil#\hfil &
  \hfil#\hfil & \hfil#\hfil & \hfil#\hfil & \hfil# & \hfil#\hfil &
\hfil# \cr
 \multispan{9} \hfil {Table 1 (continued) }
\hfil \cr
 \noalign{ \vskip 12pt } \noalign{  \hrule
 height 1pt } \noalign{ \vskip 8pt } HD & B$-$V & $(b-y)$ & $ m_1 $ &
 $c_1$ & $\beta$ & [Me/H] & rejected? & W vel \cr
 \noalign{ \vskip 12pt }
 \noalign{ \hrule height 1pt }
 \noalign{ \vskip 8pt }
   64606 & 0.73  & 0.450   & 0.221   & 0.217   & \dots   & $-$0.76   & no &
$-$5 \cr
   65583 & 0.71  & 0.449   & 0.234   & 0.229    & 2.548   & $-$0.61   & no &
$-$35 \cr
   66751 & 0.58  & 0.372   & 0.151   & 0.265    & 2.585   & $-$0.52   & no & 9
\cr
   67458 & 0.60  & 0.375   & 0.189   & 0.302   & \dots   & $-$0.07  &
yes & \dots  \cr
   68456 & 0.43  & 0.290   & 0.143   & 0.453   & \dots   & $-$0.31   & yes &
\dots \cr
   69830 & 0.76  & 0.458   & 0.297   & 0.307   & \dots   &
\phantom{$-$}0.01  & no & \dots  \cr
   73524 & 0.60  & 0.376   & 0.202   & 0.376   & \dots   & \phantom{$-$}0.17
& yes & \dots \cr
   77137 & 0.69   & 0.438   & 0.234   & 0.362   & \dots   &
\phantom{$-$}0.02 & no  & \dots \cr
   78612 & 0.61  & 0.384   & 0.185   & 0.313   & \dots   & $-$0.17  & yes &
\dots \cr
   88261 & 0.60  & 0.376   & 0.168   & 0.295   & \dots   & $-$0.34  & no &
\dots \cr
   88697 & 0.49  & 0.326   & 0.163   & 0.437   & \dots   & $-$0.12    & yes &
\dots \cr
   88725 & 0.60  & 0.394   & 0.166   & 0.256    & 2.577   & $-$0.59  & no &
$-$24 \cr
   88742 & 0.59  & 0.379   & 0.181   & 0.334   & \dots   & $-$0.14   & yes &
\dots \cr
   89707 & 0.55  & 0.360   & 0.148   & 0.307    & 2.603   & $-$0.48  & no & 54
\cr
   94444 & 0.52  & 0.343   & 0.135   & 0.316    & 2.605   & $-$0.55   & no &
$-$8 \cr
   94518 & 0.60  & 0.378   & 0.179   & 0.288   & \dots   & $-$0.26  & no &
\dots  \cr
   96700 & 0.60  & 0.389   & 0.176   & 0.319   & \dots   & $-$0.29  & no &
\dots \cr
   97343 & 0.77  & 0.458   & 0.300   & 0.313   & \dots   & \phantom{$-$}0.04 &
no   \cr
  100004 & 0.41  & 0.279   & 0.126   & 0.466    & 2.674   & $-$0.57  &
yes & $-$2  \cr
  102438 & 0.68  & 0.426   & 0.220   & 0.273   & \dots   & $-$0.31  & no &
\dots \cr
  104304 & 0.77  & 0.464   & 0.312   & 0.345    & 2.583   &
\phantom{$-$}0.17 & no & \dots  \cr
  110010 & 0.59  & 0.395   & 0.228   & 0.387    & 2.609   & \phantom{$-$}0.31
& yes & \dots \cr
    \noalign{ \vskip 8pt\hrule height 1pt
  \vskip 8pt }
}}$$

   \vfill\eject
  $$ \vbox{ \halign to \hsize { \hfil#\hfil & \hfil#\hfil & \hfil#\hfil &
  \hfil#\hfil & \hfil#\hfil & \hfil#\hfil & \hfil# & \hfil#\hfil &
\hfil# \cr
 \multispan{9} \hfil {Table 1 (continued) }
\hfil \cr
 \noalign{ \vskip 12pt } \noalign{  \hrule
 height 1pt } \noalign{ \vskip 8pt } HD & B$-$V & $(b-y)$ & $ m_1 $ &
 $c_1$ & $\beta$ & [Me/H] & rejected? & W vel  \cr
 \noalign{ \vskip 12pt }
 \noalign{ \hrule height 1pt }
 \noalign{ \vskip 8pt }
  112164 & 0.64  & 0.392   & 0.209   & 0.458   & \dots   &
\phantom{$-$}0.16 & yes & \dots  \cr
  114174 & 0.67  & 0.418   & 0.233   & 0.335    & 2.589   & \phantom{$-$}0.07
& no & \dots \cr
  116459 & 0.50  & 0.339   & 0.165   & 0.399   & \dots   & $-$0.15   & yes &
\dots \cr
  117635 & 0.78  & 0.474   & 0.293   & 0.240   & \dots   & $-$0.47  & no &
$-$19 \cr
  117939 & 0.67  & 0.409   & 0.208   & 0.310   & \dots   & $-$0.14  & no &
\dots \cr
  120690 & 0.69  & 0.434   & 0.236   & 0.301   & \dots   & $-$0.14   & no &
\dots \cr
  122742 & 0.74  & 0.451   & 0.272   & 0.319   & \dots   & \phantom{$-$}0.02  &
no & \dots \cr
  122862 & 0.58  & 0.366   & 0.185   & 0.359   & \dots   & $-$0.05  & yes &
\dots \cr
  124292 & 0.74  & 0.446   & 0.269   & 0.295   & \dots   & $-$0.08  & no &
\dots \cr
  129333 & 0.61  & 0.408   & 0.202   & 0.301   & \dots   & $-$0.21   & no &
\dots \cr
  130948 & 0.56  & 0.376   & 0.192   & 0.324   & \dots   & $-$0.03  & yes &
\dots \cr
  131511 & 0.84  & 0.499   & 0.371   & 0.294   & \dots   & $-$0.13  & no &
\dots \cr
  133002 & 0.68  & 0.442   & 0.192   & 0.345    & 2.567   & $-$0.41   & no &
$-$18 \cr
  134483 & 0.50  & 0.322   & 0.165   & 0.395   & \dots   & $-$0.07 & yes &
\dots  \cr
  136352 & 0.65  & 0.403   & 0.195   & 0.294    & 2.592   & $-$0.26  & no &
\dots \cr
  143291 & 0.77  & 0.466   & 0.284   & 0.252   & \dots   & $-$0.38   & no &
\dots \cr
  144287 & 0.77  & 0.467   & 0.287   & 0.316   & \dots   & $-$0.03  & no &
\dots \cr
  145417 & 0.83  & 0.511   & 0.276   & 0.168   & \dots   & $-$1.28  & no &
\dots \cr
  145675 & 0.88  & 0.528   & 0.393   & 0.409    & 2.575   &
\phantom{$-$}0.11 & no & \dots  \cr
  145809 & 0.61  & 0.396   & 0.187   & 0.328   & \dots   & $-$0.20  & no &
\dots \cr
  146775 & 0.61  & 0.379   & 0.211   & 0.318   & \dots   &
\phantom{$-$}0.11 & yes & \dots  \cr
  150706 & 0.62  & 0.389   & 0.188   & 0.312   & \dots   & $-$0.18  & no &
\dots \cr
    \noalign{ \vskip 8pt\hrule height 1pt
  \vskip 8pt }
}}$$

   \vfill\eject
  $$ \vbox{ \halign to \hsize { \hfil#\hfil & \hfil#\hfil & \hfil#\hfil &
  \hfil#\hfil & \hfil#\hfil & \hfil#\hfil & \hfil# & \hfil#\hfil &
\hfil# \cr
 \multispan{9} \hfil {Table 1 (continued) }
\hfil \cr
 \noalign{ \vskip 12pt } \noalign{  \hrule
 height 1pt } \noalign{ \vskip 8pt } HD & B$-$V & $(b-y)$ & $ m_1 $ &
 $c_1$ & $\beta$ & [Me/H] & rejected? & W vel \cr
 \noalign{ \vskip 12pt }
 \noalign{ \hrule height 1pt }
 \noalign{ \vskip 8pt }
  152391 & 0.76  & 0.457   & 0.294   & 0.290   & \dots   & $-$0.08   & no &
\dots \cr
  153597 & 0.48  & 0.327   & 0.154   & 0.364    & 2.627   & $-$0.24  & yes &
\dots \cr
  153631 & 0.58  & 0.386   & 0.189   & 0.339   & \dots   & $-$0.09 & yes &
\dots \cr
  154345 & 0.73  & 0.448   & 0.273   & 0.286    & 2.565   & $-$0.12  & no &
\dots \cr
  154577 & 0.89  & 0.519   & 0.389   & 0.227   & \dots   & $-$0.51  & no & 17
\cr
  155918 & 0.60  & 0.385   & 0.153   & 0.269    & 2.590   & $-$0.62   & no & 70
\cr
  158633 & 0.76  & 0.460   & 0.286   & 0.244    & 2.550   & $-$0.39   & no &
\dots \cr
  159222 & 0.65  & 0.404   & 0.219   & 0.364    & 2.595   &
\phantom{$-$}0.13 & yes & \dots  \cr
  161612 & 0.71  & 0.439   & 0.259   & 0.359   & \dots   & \phantom{$-$}0.17  &
no & \dots \cr
  162521 & 0.45  & 0.292   & 0.160   & 0.419   & \dots   & $-$0.06  & yes &
\dots \cr
  165401 & 0.62  & 0.391   & 0.170   & 0.286    & 2.581   & $-$0.44  & no &
$-$38 \cr
  166620 & 0.87  & 0.516   & 0.407   & 0.294    & 2.540   & $-$0.16   & no &
\dots \cr
  167954 & 0.54  & 0.335   & 0.180   & 0.412   & \dots   & \phantom{$-$}0.08 &
yes & \dots \cr
  171665 & 0.69  & 0.420   & 0.230   & 0.309   & \dots   & $-$0.05 &
no & \dots  \cr
  172051 & 0.68  & 0.415   & 0.225   & 0.267    & 2.569   & $-$0.23  & no &
\dots \cr
  176377 & 0.58  & 0.385   & 0.179   & 0.292    & 2.588   & $-$0.30  & no &
\dots \cr
  177565 & 0.71  & 0.431   & 0.256   & 0.326    & 2.584   & \phantom{$-$}0.09
& no & \dots \cr
  182488 & 0.81  & 0.481   & 0.348   & 0.331    & 2.567   & \phantom{$-$}0.08
& no & \dots \cr
  189567 & 0.65  & 0.399   & 0.199   & 0.298   & \dots   & $-$0.19 & no & \dots
\cr
  190248 & 0.76  & 0.472   & 0.278   & 0.398    & 2.603   &
\phantom{$-$}0.16 & no & \dots  \cr
  192310 & 0.88  & 0.521   & 0.428   & 0.295   & \dots   & $-$0.16  & no &
\dots \cr
  193307 & 0.55  & 0.363   & 0.148   & 0.349    & 2.604   & $-$0.49   & no & 21
\cr
    \noalign{ \vskip 8pt\hrule height 1pt
  \vskip 8pt }
}}$$

   \vfill\eject
  $$ \vbox{ \halign to \hsize { \hfil#\hfil & \hfil#\hfil & \hfil#\hfil &
  \hfil#\hfil & \hfil#\hfil & \hfil#\hfil & \hfil# & \hfil#\hfil &
\hfil# \cr
 \multispan{9} \hfil {Table 1 (continued) }
\hfil \cr
 \noalign{ \vskip 12pt } \noalign{  \hrule
 height 1pt } \noalign{ \vskip 8pt } HD & B$-$V & $(b-y)$ & $ m_1 $ &
 $c_1$ & $\beta$ & [Me/H] & rejected? & W vel \cr
 \noalign{ \vskip 12pt }
 \noalign{ \hrule height 1pt }
 \noalign{ \vskip 8pt }
  193664 & 0.58  & 0.386   & 0.184   & 0.315    & 2.593   & $-$0.19 & no &
\dots \cr
  194640 & 0.73  & 0.440   & 0.274   & 0.301   & \dots   & \phantom{$-$}0.01 &
no & \dots \cr
  195987 & 0.79  & 0.481   & 0.299   & 0.270   & \dots   & $-$0.31  & no &
\dots \cr
  196761 & 0.72  & 0.441   & 0.259   & 0.258   & \dots   & $-$0.28  & no &
\dots \cr
  197214 & 0.67  & 0.423   & 0.217   & 0.250   & \dots   & $-$0.42  & no & 11
\cr
  199288 & 0.58  & 0.384   & 0.148   & 0.265    & 2.575   & $-$0.68   & no & 43
\cr
  199620 & 0.84  & 0.527   & 0.294   & 0.391   & \dots   & $-$0.10  & no &
\dots \cr
  202457 & 0.68  & 0.433   & 0.220   & 0.370   & \dots   & $-$0.04   & no &
\dots \cr
  205905 & 0.62  & 0.388   & 0.219   & 0.329   & \dots   & \phantom{$-$}0.15 &
yes & \dots \cr
  210277 & 0.75  & 0.472   & 0.275   & 0.382   & \dots   & \phantom{$-$}0.11  &
no & \dots \cr
  210918 & 0.65  & 0.398   & 0.221   & 0.320   & \dots   & \phantom{$-$}0.07  &
yes & \dots \cr
  213628 & 0.72  & 0.441   & 0.266   & 0.315   & \dots   & \phantom{$-$}0.03  &
no & \dots \cr
  213941 & 0.66  & 0.418   & 0.193   & 0.270   & \dots   & $-$0.46  & no &
$-$69 \cr
  218209 & 0.65  & 0.419   & 0.189   & 0.258    & 2.568   & $-$0.54
& no & $-$14 \cr
  220339 & 0.89  & 0.515   & 0.414   & 0.251   & \dots   & $-$0.34   & no &
\dots \cr
  221613 & 0.58  & 0.394   & 0.183   & 0.286   & \dots   & $-$0.33  & no &
\dots \cr
  223731 & 0.43  & 0.293   & 0.138   & 0.428   & \dots   & $-$0.38   & yes &
\dots \cr
 \noalign{ \vskip 8pt\hrule height 1pt \vskip 1pt \hrule height 1pt
  \vskip 8pt }
}}$$

   \vfill\eject
  $$ \vbox{ \halign to \hsize { \hfil # & \hfil # \cr
 \multispan{2} \hfil {Table 2 : { Adjusted Turn-off $(b-y)$ Colors}}
\hfil \cr
 \noalign{ \vskip 12pt }
 \noalign{ \hrule height 1pt \vskip 1pt \hrule height 1pt }
 \noalign{ \vskip 8pt }
    [Fe/H] & $(b-y)$ \cr
 \noalign{ \vskip 8pt\hrule height 1pt
  \vskip 8pt }
 $-2.27 $ & 0.241 \cr
 $-1.77 $ & 0.260 \cr
 $-1.27 $ & 0.288 \cr
 $-0.79 $ & 0.323 \cr
 $-0.49 $ & 0.348 \cr
 $0.0{\phantom{0}}$ & 0.406 \cr
 $+0.50^{a}$  & 0.513 \cr
 $+0.50^b$  & 0.419 \cr
 \noalign{ \vskip 8pt\hrule height 1pt \vskip 1pt \hrule height 1pt
  \vskip 8pt }
}}$$

\noindent
Notes to Table 2 :
\item{$(a)$} Interpolation in Table 1 of vandenBerg and Bell
\item{$(b)$} Predicted from Edvardsson \etal\

\vfill\eject
\hsize=8.5truein
  $$ \vbox{ \halign to \hsize { #\hfil & \hfil#\hfil & \hfil#\hfil &
  \hfil#\hfil & \hfil#\hfil & \hfil#\hfil & \hfil#\hfil & \hfil#\hfil \cr
 \multispan{8} \hfil {Table 3: Abundance Distribution Functions for Stellar
Populations } \hfil \cr
 \noalign{ \vskip 12pt }
\noalign{  \hrule  height 1pt \vskip 1pt \hrule height 1pt }
\noalign{ \vskip 8pt }
[Fe/H] & Field Halo & Outer Bulge & Young Disk & Thin Disk  & Thick
Disk  & Thin + Thick \cr
 & & & & Volume & Volume & Column \cr
 & Note 1 & Note 2 & Note 3 & Note 4 & Note 4 & Note 5 \cr
 \noalign{ \vskip 8pt }
 \noalign{ \hrule height 1pt }
 \noalign{ \vskip 8pt }
$+0.3$ & $-$ & $-$ & $-$ & $-$ & $-$ & $-$ \cr
$+0.1$ & $-$ & $-$ & 0.65 & 0.55 &  $-$ & 0.55 \cr
$-0.1$ & 0.00 & 0.99 & 1.00 & 1.00 & 0.00: & 1.00 \cr
$-0.3$ & 0.15 & 1.00 & 0.35 & 0.52 & 0.67 & 0.72 \cr
$-0.5$ & 0.19 & 0.92 & 0.06 & 0.45 & 1.00 & 0.75 \cr
$-0.7$ & 0.23 & 0.75 & 0.00 & 0.06 & 1.00 & 0.36 \cr
$-0.9$ & 0.27 & 0.61 & 0.00 & 0.00 & 1.00 & 0.30 \cr
$-1.1$ & 0.37 & 0.42 & 0.00 & 0.00 & 0.10: & 0.03 \cr
$-1.3$ & 0.53 & 0.32 & 0.00 & 0.00 & 0.10: & 0.03: \cr
$-1.5$ & 0.85 & 0.20 & 0.00 & 0.00 & 0.00 & 0.00 \cr
$-1.7$ & 0.94 & 0.14 & 0.00 & 0.00 & 0.00 & 0.00 \cr
$-1.9$ & 0.65 & 0.08 &  0.00 & 0.00 & 0.00 & 0.00 \cr
$-2.1$ & 0.48 & 0.08: & 0.00 & 0.00 & 0.00 & 0.00 \cr
$-2.3$ & 0.34 & 0.03: & 0.00 & 0.00 & 0.00 & 0.00 \cr
$-2.5$ & 0.24 & 0.03: & 0.00 & 0.00 & 0.00 & 0.00 \cr
\noalign{ \vskip 8pt\hrule height 1pt \vskip 1pt \hrule height 1pt
  \vskip 8pt }
}}$$
\vfill\eject
\hsize=14truecm
\noindent
Notes to Table 3:

Note 1: data from Laird {\it et al } (1988).

{Note 2:} data from Ibata and Gilmore (1995).

{Note 3:} Gliese pruned stars, this paper.

{Note 4:} This paper, \S 3.

{Note 5:} This paper. Weights (from Figure 9) are, for
isothermal samples with vertical dispersions 42km/s and 19km/s, in the
ratio 3.09.

\bye